\documentclass[pre,aps,superscriptaddress,showpacs]{revtex4}

\def \beq{\begin{equation}}         \def \eeq{\end{equation}}
\def \beqa{\begin{eqnarray}}        \def \eeqa{\end{eqnarray}}
\def \bea{\begin{array}}        \def \eea{\end{array}}

\usepackage{amsmath}
\usepackage{epsfig}
\usepackage[dvips]{color}

\begin{document}
\title{A generalized integral fluctuation theorem for general jump processes }
\author{Fei Liu}
\email[Email address:]{liufei@tsinghua.edu.cn} \affiliation{Center
for Advanced Study, Tsinghua University, Beijing, 100084, China}
\author{Yu-Pin Luo}
\affiliation{Department of Electronic Engineering, National
Formosa University, Yunlin County 632, Taiwan}
\author{Ming-Chang Huang}
\affiliation{Department of Physics and center for Nonlinear and
Complex systems, Chung-Yuan Christian University, Chungli, 32023
Taiwan}
\author{Zhong-can Ou-Yang}
\affiliation{Center for Advanced Study, Tsinghua University,
Beijing, 100084, China} \affiliation{Institute of Theoretical
Physics, The Chinese Academy of Sciences, P.O.Box 2735 Beijing
100080, China}
\date{\today}

\begin{abstract}
{Using the Feynman-Kac and Cameron-Martin-Girsanov formulas, we
obtain a generalized integral fluctuation theorem (GIFT) for
discrete jump processes by constructing a time-invariable inner
product. The existing discrete IFTs can be derived as its specific
cases. A connection between our approach and the conventional
time-reversal method is also established. Different from the
latter approach that were extensively employed in existing
literature, our approach can naturally bring out the definition of
a time-reversal for a Markovian stochastic system. Intriguingly,
we find the robust GIFT usually does not result into a detailed
fluctuation theorem. }
\end{abstract}
 \maketitle

\maketitle

\section{Introduction}
One of important progresses in nonequilibrium statistic physics in
the past two decades is the discovery of a various of fluctuation
theorems. They are thought of to be a nonperturbative extension of
the fluctuation-dissipation theorems in near-equilibrium region to
far-from equilibrium region. According to their mathematical
expressions, these theorems are loosely divided into two types.
One is called the integral fluctuation theorems
(IFT)~\cite{JarzynskiPRE97,JarzynskiPRL97,
Crooks99,Crooks00,Maes,SeifertPRL05,HatanoSasa,Speck,Schmiedl,QianJPC,Harris},
and the other is called the detailed fluctuation theorems
(DFT)~\cite{Evans,Gallavotti,Kurchan,Lebowitz,Crooks00}. The
former follows a unified expression
\begin{eqnarray}
\langle \exp[-{\cal A}]\rangle=1, \label{IFT}
\end{eqnarray}
where $\cal{A}$ is a functional of a stochastic trajectory of a
concerned stochastic system, and angular brackets denote an
average over the ensemble of the trajectories that start from an
given initial distribution. For instance, $\cal{A}$ may be the
dissipated work along a trajectory and eq.~(\ref{IFT}) is the
celebrated Jarzynski equality
(JE)~\cite{JarzynskiPRE97,JarzynskiPRL97}.

Due to the insightful work of Hummer and Szabo~\cite{Hummer01}, we
now know that these IFTs have an intimate connection with the
famous Feynman-Kac formula (FK)~\cite{Feynman,Kac} in the
stochastic theory of diffusion processes~\cite{Stroock}. Recently,
several works including us reinvestigated this issue from
mathematic generalization and
rigors~\cite{QianJPC,Ge,Chetrite,Ao,LiuF}. One of findings is that
the application of the FK formula in proving the IFTs is based on
a construction of a time-reversed process of a diffusion
process~\cite{Chetrite,LiuF}. Because the definition of a
time-reversal has some certain arbitrariness~\cite{Chetrite}, we
have obtained a generalized IFT (GIFT) by constructing a
time-invariable integral and employing the FK and
Cameron-Martin-Girsanov (CMG) formulas~\cite{Cameron,Girsanov}
simultaneously, and the several
IFTs~\cite{JarzynskiPRE97,SeifertPRL05,HatanoSasa,QianJPC} were
specific cases of the GIFT~\cite{LiuF}. We should emphasize that
all of the works were concerning with continuous diffusion
processes described by Fokker-Planck (FK) equation.

In addition to continuous case, there are still another kind of
stochastic jump processes described by Markovian discrete master
equations. In many practical physical systems, a description of
discrete jump process is more satisfactory than a description
using continuous diffusion process, e.g., the systems only
involving few individual objects~\cite{Gardiner}. One may
naturally think of that there exists a GIFT in discrete version,
and the discrete IFTs in
literature~\cite{SeifertPRL05,Harris,SeifertJPA04} are specific
cases of it.  At a first sight, this effort seems trivial since a
continuous diffusion process can be always discretized to a
discrete jump process. However, In addition that one hardly
ensures that the ``discrete" GIFT achieved in this way is really
exact, we know that a jump process is not always equivalent to a
discretization of a certain continuous process~\cite{Gardiner}.
Additionally, to our knowledge, fewer works have formally studied
the IFTs for general jump processes employing the FK and CMG
formulas, though several authors have mentioned this
possibility~\cite{QianJP,GeJMP} earlier. Therefore, in our opinion
a rigorous derivation of an exact GIFT for discrete jump processes
is essential and meaningful. In this work we present this effort.
Because we focus on the general Markovian jump processes, fewer
physics are mentioned here. The detailed discussions about the
specific IFTs in previous literature~\cite{Harris} should make it
up.

\section{Generalized integral fluctuations for jump processes}
We start with a Markovian jumping process described by a discrete
master equation
\begin{eqnarray}
\frac{dp_n(t)}{dt}=\left[{\textbf H}(t){\textbf p}(t)\right]_n,
\label{forwardeq}
\end{eqnarray}
where the $N$-dimension column vector ${\textbf
p}(t)=(p_1,\cdots,p_N)^{\rm T}$ is the probabilities of the system
at individual states at time $t$ (the state index $n$ may be a
vector), the matrix element of the time-dependent (or
time-independent) rate ${\textbf H}_{mn}>0$ ($m\neq n$) and
${\textbf H}_{nn}=-\sum_{m\neq n}{\textbf H}_{mn}$. Given a
normalized positive column vector ${\textbf
f}(t)=(f_1,\cdots,f_N)^{\rm T}$ and a $N\times N$ matrix ${\textbf
A}$ that satisfies conditions $f_n{\textbf H_{mn}+\textbf
A}_{mn}>0$ ($m\neq n$) and ${\textbf A}_{nn}=-\sum_{m\neq
n}{\textbf A}_{mn}$, we state that an inner product ${\textbf
f}^{\rm T}(t'){\textbf v}(t')$ is time-invariable if the column
vector $\textbf v(t')=(v_1,\cdots,v_N)^{\rm T}$ satisfies
\begin{eqnarray}
\frac{dv_n(t')}{dt'}=-\left[{\textbf H}^{\rm T}{\textbf
v}\right]_n-f_n^{-1}\left[\partial_{t'}{ \textbf f}-{\textbf H}{
\textbf f}\right]_nv_n+f_n^{-1} \left[\left({ \textbf A} \textbf
1\right)_nv_n-\left({\textbf A}^{\rm T} {\bf\text
v}\right)_n\right], \label{backwardeq}
\end{eqnarray}
where the final condition of $v_n(t)$ is $q_n$ ($t'$$<$$t$), and
the column vector ${\textbf 1}=(1,\cdots,1)^{\rm T}$. This is
easily proved by noting a time differential $d_{t'}\left[{\textbf
f}^{\rm T}(t'){\textbf v}(t')\right]= d_{t'}({\textbf f}^{\rm
T}){\textbf v}+{\textbf f}^{\rm T}d_{t'}({\textbf v})$ and the
transpose property of a matrix. Employing the Feynman-Kac and
Cameron-Martin-Girsanov formulas for jump processes (a simple
derivation about the latter see the Appendix I),
eq.~(\ref{backwardeq}) has a stochastic representation given by
\begin{eqnarray}
v_n(t')=E^{n,t'}\left[e^{-{\cal J}[{\textbf x},\textbf f,\textbf
A]} q_{s(t)}\right] \label{stochrep}
\end{eqnarray}
and
\begin{eqnarray}
{\cal J}[{\textbf x},\textbf f,\textbf A]=\int_{t'}^tf_{\textbf
x(\tau)}^{-1}\left[-\partial_{\tau}\textbf f+\textbf H\textbf
f+\textbf A\textbf 1\right]_{{\textbf
x}(\tau)}d\tau-\int_{t'}^tf_{{\textbf x}(\tau)}^{-1}\textbf
A_{{\textbf x}(\tau){\textbf x}(\tau)}d\tau
-\sum_{i=1}^k\ln\left[1+\frac{{\textbf A}_{{\textbf
x}(t_i^+){\textbf x}(t_{i}^-)}(t_i)}{f_{{\textbf
x}(t_i^-)}(t_i)\textbf H_{{\textbf x}(t_i^+){\textbf
x}(t_{i}^-)}(t_i)} \right],\label{functional}
\end{eqnarray}
the expectation $E^{n,t'}$ is over all trajectories $\textbf x$
generated from eq.(\ref{forwardeq}) with fixed initial state $n$
at time $t'$, $\textbf x(t')$ is the discrete state at time $t'$,
$\textbf x(t_i^{-})$ and $\textbf x(t_i^{+})$ represent the states
just before and after a jump occurs at time $t_i$, respectively,
and we assumed the jumps occur $k$ times for a process. The
readers are reminded that the first and last two terms of the
functional are the consequences of the FK and GCM formulas,
respectively. We see that the last term is significantly different
from that in the continuous processes [eq. (11) in
ref.~\cite{LiuF}]. Combining the stochastic representation and the
time-invariable quantity and choosing $t'=0$, we obtain the exact
discrete GIFT for a jump process,
\begin{eqnarray}
\sum_{m=1}^N f_m(0) E^{m,0}\left\{e^{-J[{\textbf x},\textbf
f,\textbf A]} q_{{\textbf x}(t)}]\right\}={\textbf f}^{\rm T}(t)
{\textbf q} \label{GIFT}
\end{eqnarray}
Particulary, the right hand side of the equation become $1$ if
$\textbf q=1$.

\section{Relationship between the GIFT and existing IFTs for jump processes}
\label{secIII} The abstract eq.~(\ref{GIFT}) includes several
discrete IFTs in literature. First we investigate the case in
which the discrete system has a transient steady-state solution
${\textbf H}(t) {\textbf p}^{\rm ss}(t)=0$. Choosing the matrix
$\textbf A$=0 and the vector $\textbf f(t)={\textbf p}^{\rm
ss}(t)$, eq.~(\ref{functional}) is immediately simplified into
\begin{eqnarray}
{\cal J}=-\int_0^t\partial_\tau p^{\rm ss}_{\textbf
x(\tau)}(\tau)d\tau \label{JarzynskiHantanFunctional}
\end{eqnarray}
If one further thinks of ${\textbf p}^{\rm ss}$ satisfying a
time-dependent detailed balance condition ${\textbf
H}_{mn}(t)p^{ss}_n(t)={\textbf H}_{nm}(t)p^{ss}_m(t)$, the above
functional may be analogous to the dissipated work and
eq.~(\ref{GIFT}) is the discrete version of the
JE~\cite{JarzynskiPRE97,JarzynskiPRL97}. On the other hand, if
${\textbf p}^{\rm ss}$ is a transient nonequilibrium steady-state
without detailed balance, eq.~(\ref{JarzynskiHantanFunctional})
could be rewritten to
\begin{eqnarray}
{\cal J}=\ln\frac{p^{\rm ss}_{\textbf x(0)}(0)}{p^{\rm
ss}_{\textbf x(t)}(t)}+\sum_{i=1}^k\ln\frac{p^{\rm ss}_{\textbf
x(t_i^+)}(t_i)}{p^{\rm ss}_{\textbf x(t_i^-)}(t_i)}.
\label{HantanFunctional}
\end{eqnarray}
where we used the following relationship
\begin{eqnarray}
{d_t}\ln p^{\rm ss}_{\textbf x(t)}(t)=\partial_t\ln p^{\rm
ss}_{\textbf x(t)}(t)+\sum_{i=1}^{k}\delta(t-t_i)
 \ln\left[p^{\rm ss}_{\textbf x(t_i^+)}(t_i)/p^{\rm
 ss}_{\textbf x(t_i^-)}(t_i)\right].
\label{differentialrelation}
\end{eqnarray}
Then we may interpret the first term in
eq.~(\ref{HantanFunctional}) as the entropy change of system and
the second term as the ``excess" heat of the driven jump process.
Under this circumstance eq.~(\ref{GIFT}) is the discrete version
of the Hatano-Sasa equality~\cite{HatanoSasa}.

The last case is about nonvanishing $\textbf A(t)$. Choosing the
matrix element ${\textbf A}_{mn}(t)={\textbf
H}_{nm}(t)f_m(t)-{\textbf H}_{mn}(t)f_n(t)$ ($m\neq n$), or the
flux $J_{mn}(t)$ between states $m$ and $n$ for a distribution
$\textbf f(t)$. Obviously, the condition of $f_n\textbf
H_{mn}+\textbf A_{mn}$$>0$. Substituting this matrix into
eq.~(\ref{functional}), we obtain
\begin{eqnarray}
{\cal J}=-\int_0^t \partial_{\tau}\ln p_{{\textbf
x}(\tau)}(\tau)d\tau+\sum_{i=1}^{k}\ln\frac{{\textbf H}_{{\textbf
x}(t_i^-)\textbf x(t_i^+)}(t_i)p_{\textbf x(t_i^+)}(t_i)}{{\textbf
H}_{\textbf x(t_i^+)\textbf x(t_i^-)}(t_i)p_{\textbf
x(t_i^-)}(t_i)} \label{orgtotentropy}
\end{eqnarray}
To achieve obvious physical meaning of the above expression, we
employ eq.~(\ref{differentialrelation}) again and have
\begin{eqnarray} {\cal
J}=\ln\frac{f_{\textbf x(0)}(0)}{f_{\textbf
x(t)}(t)}+\sum_{i=1}^k\ln\frac{{\textbf H}_{\textbf
x(t_i^+)\textbf x(t_i^-)}(t_i)}{{\textbf H}_{\textbf
x(t_i^-)\textbf x(t_i^+)}(t_i)}. \label{totalentropy}
\end{eqnarray}
Hence, if $\textbf f(t)$ is the distribution of the system itself
satisfying the evolution eq.~(\ref{forwardeq}), the first term in
the equation is just the entropy change of the system and the
second term is interpreted as entropy change of
environment~\cite{SeifertPRL05,Harris}. In other words, the GIFT
with eq.~(\ref{totalentropy}) is about the total entropy change of
a stochastic jump process.

\section{The GIFT and time reversal for jump processes}
Like the case of continuous diffusion processes, we can connect
the time-invariable inner product to be a jump process that is
regarded to be a time-reversal of the original jump
process~\cite{LiuF}. Multiplying $f_n(t')$ and rearranging on both
sides of eq.~(\ref{backwardeq}), we have
\begin{eqnarray}
\frac{d}{dt'}\left[f_n(t')v_n(t')\right]=-\sum_{m=1}^N
f_m^{-1}\left[\textbf H_{mn}f_n+{\textbf
A}_{mn}\right]f_mv_m+f_nv_n\sum_{m=1}^Nf_n^{-1}\left[\textbf
H_{nm}f_m+{\textbf A}_{nm}\right] \label{orgtimereversal}
\end{eqnarray}
Then we define a new function $q_{\bar n}(s)=f_n(t')v_n(t')$,
where $s=t-t'$ and $\bar n$ represents an index whose components
are the same or the minus of the components of the index $n$
depending on whether they are even or odd under time reversal
($t\to -t$). We also define a new rate matrix $\overline{\textbf
H}(s)$ whose elements are
\begin{eqnarray}
\overline{\textbf H}_{{\bar n}{\bar
m}}(s)=f_m^{-1}(t')\left[\textbf H_{mn}(t')f_n(t')+\textbf
A_{mn}(t')\right] \label{ratetimereversal}
\end{eqnarray}
for $m\neq n$, and $\overline{\textbf H}_{mm}(s)=-\sum_{n\neq
m}\overline{\textbf H}_{nm}(s)$, respectively. Then
eq.~(\ref{orgtimereversal}) is rewritten as
\begin{eqnarray}
\frac{dq_{\bar n}(s)}{ds}=[\overline{\textbf H}(s)\textbf
q(s)]_{\bar n}. \label{timereversal}
\end{eqnarray}
Because of the variable $s=t-t'$, we interpret $\overline{\textbf
H}(t)$ to be a time-reversal of the original $\textbf H(t)$.
Equation~(\ref{timereversal}) directly presents the reason of the
time-invariable inner product $\textbf f^{\rm T}(t')\textbf v(t')$
that equals $\textbf 1^{\rm T}\textbf q(s)$; the latter is a
constant due to probability conservation.

The generalized time-reversal~(\ref{ratetimereversal}) includes
several types of time-reversal in
literature~\cite{HatanoSasa,Chernyak,Harris}. For convenience, we
only consider even components only in the state-index $n$. First,
if the matrix $\textbf A=0$ and $\textbf f(t')=\textbf p^{\rm
ss}(t')$ satisfying the detailed balance condition, the
time-reversed rate matrix $\overline{\textbf H}(t')={\textbf
H}(s)$ simply. The process determined by this rate matrix was
termed backward process~\cite{Harris} (or a reversed protocol in
Ref.~\cite{Chernyak}). In contrast, if $\textbf p^{\rm ss}(t')$ is
transient nonequilibrium steady-state, a process determined by
$\overline{\textbf H}_{mn}(t')=f_n(s){\textbf H}_{nm}(s)/f_m(s)$
was termed an adjoint process~\cite{Harris} (or the current
reversal in Ref.~\cite{Chetrite}). Intriguingly, if we choose
$\textbf A_{mn}(s)$ to be the flux $J_{mn}(t)$ between the states
$m$ and $n$ for a distribution $\textbf f(s)$, we reobtain
$\overline{\textbf H}(t')={\textbf H}(s)$ that is the same with
case of the detailed balance condition. Considering that these
choices of $\textbf f$ and $\textbf A$ here are corresponding to
those in Sec.~\ref{secIII}, respectively, we see that the JE and
the IFT of the total entropy have the same physical origin. It is
expected in physics since a realization of a reversed protocol is
usually possible and does not depend on whether the system
satisfies detailed balance condition. We should point out that one
may construct infinite time-reversals, because $\textbf f$ and
$\textbf A$ are almost completely arbitrary, e.g., $\textbf
A_{mn}(s)=\alpha J_{mn}(s)$ and $0\leq\alpha\leq 1$. Before ending
this section, we give two comments about the relationship $q_{\bar
n}(s)=f_n(t')v_n(t')$. First, for a time-independent $\textbf H$,
if $f_n$ is the equilibrium solution of the rate matrix,
eq.~(\ref{backwardeq}) with zero $\textbf A$ is just the backward
master equation~\cite{Gardiner}. Second, employing the
relationship repeatedly, we may obtain the detailed DFTs for the
specific vectors $\textbf f(t')$ and matrixes $\textbf A(t')$ in
Sec.~\ref{secIII} (the details see the Appendix II).

\section{Conclusion}
In this work we derived a GIFT for general jump processes. The
existing IFTs for discrete master equations are its special cases.
We see that, in form the GIFT for the jump cases is apparently
distinct from that for the continuous diffusions that we obtained
earlier~\cite{LiuF}. Additionally, we also find this robust GIFT
usually does not result into a detailed fluctuation theorem.
Compared to other approaches, the major advantage of the current
and previous works is that the time-reversal can come out
automatically during the constructions of the time-invariable
integral or the inner product, which should be direct and obvious,
at least from point of view of us. Of course, A limit of our two
works is that we did not show some applications of the two GIFTs
in concrete physical systems. We hope that this point would be
remedied in near future. \\

{\noindent This work was supported in part by Tsinghua Basic
Research Foundation and by the National Science Foundation of
China under Grant No. 10704045 and No. 10547002.}

\appendix
\section*{Appendix I: The Cameron-Martin-Girsanov formula for jump processes}
Compared to the CMG formula for continuous diffusion processes,
little literature discussed the CMG formula for discrete jump
processes. For the convenience of the readers, we give a simple
derivation of the formula here. Given a master equation with rate
matrix ${\textbf H}$. The probability observing a trajectory
${\textbf x}(\cdot)$ which starts state $n_1$ at time $t_0=0$,
jumps at time $t_1$ to state $n_2$,$\cdots$, finally jumps at time
$t_k$ to $n_{k+1}$ and stay till time $t_{k+1}=t$ is
\begin{eqnarray}
{\rm prob}[{\textbf
x}(\cdot)]=&&\prod_{i=1}^{k}\exp\left[\int_{t_{i-1}}^{t_i}{\textbf
H}_{\textbf x(t_i^-)\textbf x(t_i^-)}(\tau)d\tau\right]{\textbf
H}_{\textbf x(t_i^-)\textbf x(t_i^+)}\times
\exp\left[\int_{t_{k}}^{t}{\textbf
H}_{n_{k+1}n_{k+1}}(\tau)d\tau\right],
\end{eqnarray}
where $\textbf x(t_i^{-})=n_i$ and $\textbf x(t_i^{+})=n_{i+1}$
($i=1,\cdots,k$). Assuming that there is another master equation
with a different rate matrix ${\textbf H'}={\textbf H}+{\textbf
A}$, where the matrix elements of ${\textbf A}$ may be negative.
Then the ration of the probabilities observing the same trajectory
in these two equations is simply
\begin{eqnarray}
\frac{{\rm prob}'[{\textbf x}(\cdot)]}{{\rm prob}[{\textbf
x}(\cdot)]}=e^{-Q[{\textbf x}(\cdot)]}. \label{CGMformula}
\end{eqnarray}
where
\begin{eqnarray}
Q[{\textbf x}(\cdot)]=-\int_{0}^{t}{\textbf A}_{\textbf
x(\tau)\textbf
x({\tau})}(\tau)d\tau-\sum_{i=1}^{k}\ln\left(1+\frac{ {\textbf
A}_{\textbf x(t_i^-)\textbf x(t_i^+)}}{{\textbf H}_{\textbf
x(t_i^-)\textbf x(t_i^+)}}\right)
\end{eqnarray}
Obviously, eq.~(\ref{CGMformula}) results into a IFT
\begin{eqnarray}
\langle e^{-Q[{\textbf x}(\cdot)]}\rangle =1,
\end{eqnarray}
where the average is over an ensemble of trajectories generated
from the stochastic system with the rate matrix $\textbf H$ and
with any initial distribution. Choosing a specific
\begin{eqnarray}
{\textbf A}_{mn}(t)=p^{\rm ss}_n(t)^{-1}\left[{\textbf
H}_{nm}(t)p^{\rm ss}_m(t)-{\textbf H}_{mn}(t)p^{\rm
ss}_n(t)\right]{\text (m\neq n)}
\end{eqnarray}
and ${\textbf A}_{nn}=-\sum_{m\neq n}{\textbf A}_{mn}(t)=0$, we
obtain the IFT of the house-keeping heat~\cite{Speck,Harris} in
discrete version, where
\begin{eqnarray}
Q_{\rm hk}[{\textbf x}(\cdot)]=\sum_{i=1}^{k}\frac{{\textbf
H}_{\textbf x(t_i^-)\textbf x(t_i^+)}(t_i)p^{\rm ss}_{\textbf
x(t_i^+)}(t_i)}{{\textbf H}_{\textbf x(t_i^+)\textbf x(t_i^-)}
(t_i)p^{\rm ss}_{\textbf x(t_i^-)}(t_i)}
\end{eqnarray}
Intriguingly, replacing $p_m^{\rm ss}$ above by the real
probability distribution $p_m(t)$ of the system ${\textbf H}$, one
obtains an IFT with
\begin{eqnarray}
Q[{\textbf x}(\cdot)]=\int_0^t \partial_{\tau}\ln p_{{\textbf
x}(\tau)}(\tau)d\tau+\sum_{i=1}^{k}\frac{{\textbf H}_{\textbf
x(t_i^-)\textbf x(t_i^+)}(t_i)p_{\textbf x(t_i^+)}(t_i)}{{\textbf
H}_{\textbf x(t_i^+)\textbf x(t_i^-)}(t_i)p_{\textbf
x(t_i^-)}(t_i)} \label{psudoentropy}
\end{eqnarray}
We notice that this functional is the almost same with that of the
IFT of total entropy eq.~(\ref{orgtotentropy}) except that the
first term becomes minus here. In addition, the average of
$\langle Q[{\textbf s}(\cdot)]\rangle$ is the same with the
average of the total entropy since the first terms of
eqs.~(\ref{totalentropy}) and (\ref{orgtotentropy}) vanish. We are
not very clear whether eq.~(\ref{psudoentropy}) has new physical
interpretation.

\section*{Appendix II: the detailed fluctuation theorem}
Given the transition probability of eq.~(\ref{timereversal}) to be
$q_{\bar n}(s'|m,s)$ (0$<$$s$$<$$s'$$<$$t$), the previous
relationship implies
\begin{eqnarray}
q_{\bar n}(s'|m,s)f_{\bar
m}(t-s)=f_{n}(t-s')E^{n,t-s'}\left[e^{-{{\cal
J}(t-s',t-s)}}\delta_{{\textbf x}(t-s),\bar m}\right]
\label{Gdetailedbalance}
\end{eqnarray}
if one notices the initial condition $q_{\bar
n}(s|m,s)=\delta_{\bar n,m}$, where we use ${\cal J}(t-s',t-s)$ to
denote the functional eq.~(\ref{functional}) with the lower and
upper limits $t-s'$ and $t-s$, respectively, and $\delta$ is the
Kronecker's. Now we consider a mean of a $(k+1)$-point function
over the time-reversed system~(\ref{timereversal}),
\begin{eqnarray}
\langle{F}[{\bar {\textbf x}}(s_k),\cdots,{\bar {\textbf
x}}(s_0)]\rangle_{\rm TR}=\sum_{n_0,\cdots,n_k}
q_{n_k}(s_k|n_{k-1},s_{k-1})\cdots
q_{n_1}(s_1|n_0,s_0)q_{n_0}(s_0){F}(\bar n_k,\cdots,\bar n_0)
\end{eqnarray}
where $0$=$s_0$$<$$s_1$$<$$\cdots$$<$$s_k$$=$$t$ and $\textbf
q(s_0)$ is the initial distribution. If we choose a specific
$\textbf q_{n_0}(s_0)= f_{\bar n_0}(t-s_0)$ and employ
eq.~(\ref{Gdetailedbalance}) repeatedly, the right hand side of
the above equation becomes
\begin{eqnarray}
\langle{e^{-{\cal J}[{\textbf x},\textbf f,\textbf A]}F}[{{\textbf
x}}(t_0),\cdots,{{\textbf x}}(t_k)]\rangle=\sum_{\bar n_k}f_{\bar
n_k}(t-s_k)E^{\bar n_k,t-s_k}\left\{ e^{-{\cal
J}(0,t)}{F}[{{\textbf x}}(t-s_k),\cdots,{{\textbf
x}}(t-s_0)]\right\}.
\end{eqnarray}
Here we define $t_i=t-s_{k-i}$ and
$0$$=$$t_0$$<$$t_1$$<$$\cdots$$<$$t_k$$=$$t$. On the basis of the
above discussion, if $k\to\infty$ the function $F$ becomes a
functional ${\cal F}$ over the space of all trajectories $\textbf
x$, and we get an identity
\begin{eqnarray}
\langle{\bar{\cal F}}\rangle_{\rm TR}= \langle e^{-{{\cal
J}[{\textbf x},\textbf f,\textbf A]}}{\cal F}\rangle,
\label{GCrook}
\end{eqnarray}
where ${\cal {\bar F}}(\textbf x)={\cal F}({\bar {\textbf x}})$
and $\bar {\textbf x}$ is simply the time-reversed trajectory of
$\textbf x$. This is a generalization of Crooks'
relation~\cite{Crooks00}. Obviously, choosing ${\cal F}$ constant,
one obtains the GIFT~(\ref{GIFT}). An important following question
is whether the GIFT results into a DFT. For the specific matrixes
$\textbf A(t)$ and vectors $\textbf f(t)$ in sec.~\ref{secIII}, we
indeed obtain several DFTs
\begin{eqnarray}
P_{\rm TR}(-J)=P(J)e^{-J}, \label{DFT}
\end{eqnarray}
by choosing ${\cal F}(\textbf x)=\delta({\cal J}[{\textbf
x},\textbf f,\textbf A]-J)$, where $P(J)$ is the probability
distribution for the quantity $\cal J$ achieved from the jump
process~(\ref{forwardeq}) and $P_{\rm TR}(J)$ is the corresponding
distribution from the time-revered system~(\ref{timereversal}).
For any a pair of $\textbf A$ and $\textbf f$, eq.~(\ref{DFT})
usually does not hold.


\begin{thebibliography}{99}
\bibitem{JarzynskiPRL97}
C. Jarzynski, Phys. Rev. Lett. {\bf 78}, 2690 (1997).

\bibitem{JarzynskiPRE97}
C. Jarzynski, Phys. Rev. E. {\bf 56}, 5018 (1997).



\bibitem{Crooks99}
G. E. Crooks, Phys. Rev. E {\bf 60}, 2721 (1999).


\bibitem{Crooks00}
G. E. Crooks, Phys. Rev. E {\bf 61}, 2361 (2000).

\bibitem{HatanoSasa}
T. Hatano and S. I. Sasa, Phys. Rev. Lett. {\bf 86}, 3463 (2001).

\bibitem{Maes}
C. Maes, Sem. Poincare, {\bf 2}, 29 (2003).

\bibitem{SeifertPRL05}
U. Seifert, Phy. Rev. Lett. {\bf 95}, 040602 (2005).

\bibitem{Speck}
T. Speck and U. Seifert, J. Phys. A {\bf 38}, L581 (2005).

\bibitem{Schmiedl}
T. Schmiedl, T. Speck, and U. Seifert, J. stat. Phys {\bf 128}, 77
(2007).

\bibitem{Harris}
R. J. Harris and G. M. Schutz, J. Stat. Mech: Theor. Exp., P07020
(2007).

\bibitem{QianJPC}
H. Qian, J. Phys. Chem. B {\bf 109}, 23624 (2005).


\bibitem{Evans}
D. J. Evans, E. G. D. Cohen, and G. P. Morriss, Phys. Rev. Lett.
{\bf 71}, 2401 (1993).

\bibitem{Gallavotti}
G. Gallavotti and E. G. D. Cohen, Phys. Rev. Lett. {\bf 74}, 2694
(1995).

\bibitem{Kurchan}
J. Kurchan, J. Phys. A, {\bf 31}, 3719 (1998).

\bibitem{Lebowitz}
J. L. Lebowitz and H. Spohn, J. Stat. Phys. {\bf 95}, 333 (1999).


\bibitem{Hummer01}
G. Hummer and A. Szabo, Proc. Proc. Natl. Acad. Sci. USA {\bf 98},
3658 (2001).




\bibitem{Feynman}
R. P. Feynman, Rev. Mod. Phys. {\bf 20} 367 (1948).

\bibitem{Kac}
M. Kac. Trans. Amer. Math. Soc. {\bf 65} 1 (1949).



\bibitem{Stroock}
D. W. Stroock and S. R. S. Varadhan, {\it Multidimensinal
Diffusion Processes}, (Springer, New York, 1979 ).


\bibitem{Ge}
H. Ge and D. Q. Jiang, J. Stat. Phys. {\bf 131}, 675 (2008).


\bibitem{Ao}
P. Ao, Commun. Theor. Phys. {\bf 49}, 1073 (2008).

\bibitem{Chetrite}
R. Chetrite and K. Gawedzki, Commun. Math. Phys. {\bf 282}, 469
(2008).


\bibitem{LiuF}
F. Liu and Zhong-can Ou-Yang, arXiv:0902.3330v2, submitted.

\bibitem{Cameron}
R. H. Cameron and W. T. Martin, Trans. Amer. Math. Soc. {\bf 75},
552 (1953).

\bibitem{Girsanov}
I. V. Girsanov, Theory of Prob. and Appl. {\bf 5}, 285 (1960).


\bibitem{Gardiner}
C. W. Gardiner, {\it Handbook of stochastic methods}, (Springer,
New York, 1983).

\bibitem{SeifertJPA04}
U. Seifert, J. Phys. A: Math. Gen. {\bf 37}, L517, (2004).


\bibitem{QianJP}
H. Qian, J. Phys.: Condens. Matter {\bf 17}, S3783 (2005).

\bibitem{GeJMP}
H. Ge and M. Qian, J. Math. Phys. {\bf 48}, 053302 (2007).

\bibitem{Chernyak}
V. Chernyak, M. Cjertkov, C. Jarzynski, J. State. Mech. P08001
(2006).


\end{thebibliography}
\end{document}